\documentclass[12pt]{article}
\usepackage{epsfig}
\usepackage{color}
\usepackage{lineno}

\begin{document}

\begin{center}

\vspace*{1.0cm}

{\Large {\bf First search for $2\varepsilon$ and
$\varepsilon\beta^+$ decay of $^{174}$Hf}}

\vskip 0.5cm

{\bf F.A.~Danevich$^{a,}$\footnote{Corresponding author. {\it
E-mail address:} danevich@kinr.kiev.ua (F.A.~Danevich).},
M.~Hult$^{b}$, D.V.~Kasperovych$^{a}$, G.P.~Kovtun$^{c,d}$,
K.V.~Kovtun$^{e}$, G.~Lutter$^{b}$, G.~Marissens$^{b}$,
O.G.~Polischuk$^{a}$, S.P.~Stetsenko$^{c}$, V.I.~Tretyak$^{a}$}

\vskip 0.3cm

$^{a}${\it Institute for Nuclear Research, 03028 Kyiv, Ukraine}

$^{b}${\it European Commission, Joint Research Centre, Retieseweg
111, 2440 Geel, Belgium }

$^{c}${\it National Scientific Center ``Kharkiv Institute of Physics
and Technology'', 61108 Kharkiv, Ukraine}

$^{d}${\it Karazin Kharkiv National University, 61022 Kharkiv,
Ukraine}

$^{e}${\it Public Enterprise ``Scientific and Technological Center
Beryllium'', 61108 Kharkiv, Ukraine}

\end{center}


\vskip 0.5cm

\begin{abstract}

The first ever search for $2\varepsilon$ and $\varepsilon\beta^+$
decay of $^{174}$Hf was realized using a high-pure sample of
hafnium (with mass 179.8 g) and the ultra low-background
HPGe-detector system located 225 m underground. After 75 days of
data taking no indication of the double beta decay transitions
could be detected but lower limits for the half-lives of the
different channels and modes of the decays were set on the level
of $\lim T_{1/2}\sim 10^{16}-10^{18}$ a.

\end{abstract}

\vskip 0.4cm


\section{INTRODUCTION}

A great interest to double beta ($2\beta$) decay, particularly to
the neutrinoless mode of the process ($0\nu2\beta$), is related to
unique possibilities to clarify properties and nature of
neutrino and weak interactions
\cite{Dolinski:2019,Barea:2012,Rodejohann:2012,Delloro:2016,Vergados:2016}, and
to test many other hypothetical scenarios of the $0\nu2\beta$
decay \cite{Dolinski:2019,Rodejohann:2012,Deppisch:2012,Bilenky:2015}.

The efforts of experimentalists are concentrated mainly on the
searches for the $0\nu$ mode of $2\beta$ decay with electrons
emission (see reviews
\cite{Delloro:2016,Tretyak:2002,Elliott:2012,Giuliani:2012,Saakyan:2013,Cremonesi:2014,Gomes:2015,Sarazin:2015}
and recent experimental works
\cite{GERDA,EXO-200,CUORE,NEMO-3,KamLAND-Zen}). However, even the
most sensitive experiments do not observe the effect and only set
half-life limits on the level of
$T^{0\nu2\beta}_{1/2}>(10^{24}-10^{26})$ a, that lead to the
restrictions on the effective Majorana mass of electron neutrino
on the level of (0.1 -- 0.7) eV, depending on the nuclei and the
nuclear matrix elements calculations. The sensitivity to the
double beta plus processes: double electron capture
($2\varepsilon$), electron capture with positron emission
($\varepsilon\beta^+$), and double positron emission ($2\beta^+$)
is much lower (we refer reader to the recent reviews
\cite{Maalampi:2013,Blaum:2020} and the references therein). At
the same time, there is a strong motivation to improve
sensitivities in studies of the double beta plus decay processes
related to a possibility to clarify possible contribution of the
right-handed currents to the $0\nu2\beta$ decay rate in case of
its observation \cite{Hirsch:1994}.

\begin{figure}[htb]
\begin{center}
 \mbox{\epsfig{figure=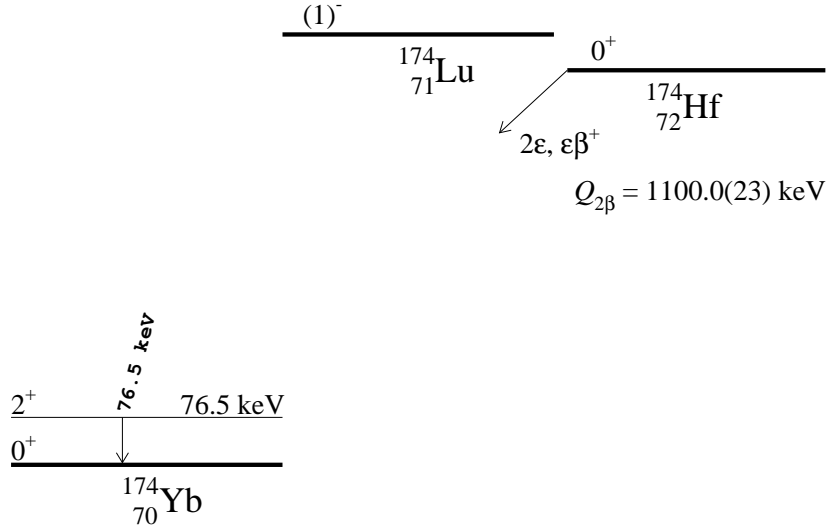,height=7.0cm}}
\caption{The simplified decay scheme of $^{174}$Hf.}
 \label{fig:174hf-scheme}
 \end{center}
 \end{figure}

The isotope $^{174}$Hf is one of the potentially $2\varepsilon$,
$\varepsilon\beta^+$ radioactive nuclides with the energy of decay
$Q_{2\beta}=1100.0(23)$ keV \cite{Wang:2017} and the isotopic
abundance $\delta=0.16(12)$\% \cite{Meija:2016}\footnote{It should
be stressed that so low isotopic abundance is typical for the
potentially double beta plus active isotopes and is one of the
practical reasons of the modest experimental sensitivity to this
kind of nuclear instability.}. A simplified expected decay scheme
of $^{174}$Hf is shown in Fig.~\ref{fig:174hf-scheme}. While the
double electron capture is possible with population of the ground
state and the first $2^+$ excited level of the daughter nuclei,
the electron capture with positron emission is allowed with the
population of the ground state only (at least, for captures from
$K$ and $L$ atomic shells, for which this process is the most
probable). All the expected decays should be accompanied by single
or multiple $\gamma$ (X-ray) quanta emission that opens a
possibility to apply HPGe $\gamma$ spectrometry to search for the
decays.

\nopagebreak

In Section \ref{sec:exp} we describe the high-purity hafnium
sample production and the experimental technique of ultra-low
background HPGe $\gamma$ spectrometry used in the present study.
The data analysis and obtained limits on the $2\varepsilon$ and
$\varepsilon\beta^+$ processes in $^{174}$Hf are reported in Sect.
\ref{sec:res}. The Conclusions section contains a summary of the
experiment and some discussion of possibilities to improve the
experimental sensitivity.

\section{Low counting experiment} \label{sec:exp}

\subsection{Hafnium sample}

A disc-shaped sample of metallic hafnium with sizes
$\oslash59.0\times 5.0$ mm (the mass of the sample was 179.8 g,
the sample contained $\approx0.29$ g of the isotope $^{174}$Hf)
was utilized in the experiment. The hafnium was produced by the
former Soviet Union industry by reduction process from hafnium
tetrafluoride with metallic calcium. Then the material was
purified by centrifugation of gaseous Hf compound to reduce
zirconium concentration that is typically the main contaminant of
hafnium, which is very hard to separate by chemical and physical
methods. Finally the material was additionally purified by double
melting in vacuum by electron beam at the National Scientific
Center ``Kharkiv Institute of Physics and Technology'' (Kharkiv,
Ukraine). The purity level of the obtained hafnium was proved by
the Laser Ablation Mass Spectrometry as
$\simeq99.8\%$\footnote{The purity level of the sample is reported
in detail in our previous work aimed to the first search for
$\alpha$ decays of naturally occurring Hf nuclides with emission
of $\gamma$ quanta \cite{Danevich:2019}.}.

\subsection{Gamma-ray spectrometry set-ups and measurements}

\begin{figure}[ht]
\begin{center}
 \mbox{\epsfig{figure=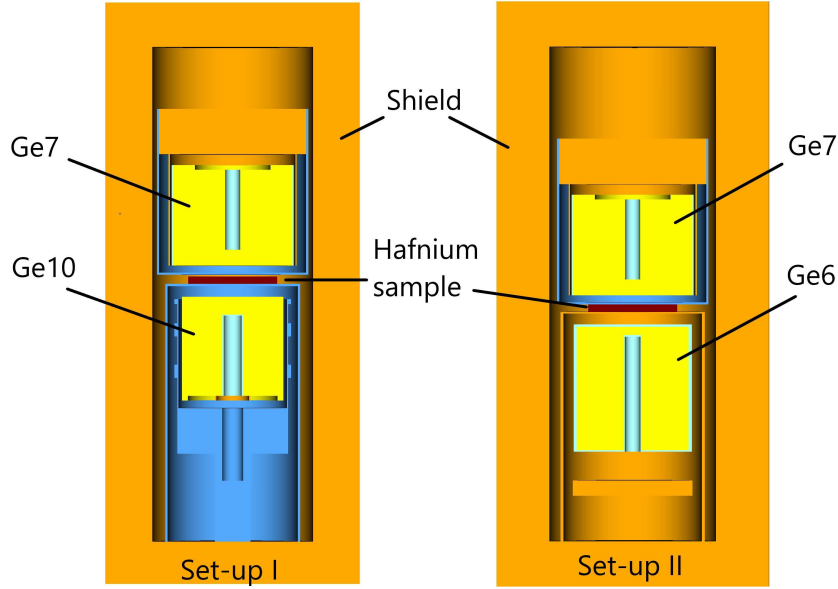,height=8.0cm}}
\caption{(Color online) Schematic view of the two low-background
set-ups with HPGe detectors and hafnium sample.}
 \label{fig:set-ups}
\end{center}
\end{figure}

The experiment was carried out with the help of two set-ups with
three HPGe detectors (named Ge6, Ge7, and Ge10) at the HADES
underground laboratory of the Joint Research Centre of European
Commission (Geel, Belgium) located at 225 m depth below the ground. A
schematic view of the both set-ups is presented in Fig.
\ref{fig:set-ups}, while the main characteristics of the detectors
are given in Table~\ref{tab:detectors}, with some more details
in \cite{Danevich:2019,Wieslander:2009,Hult:2013}.

\begin{table}[htb]
\caption{Characteristics of the HPGe-detectors used in present
experiment.}
\begin{center}
\begin{tabular}{|l|c|c|c|}
 \hline
 ~                                      & Ge6           & Ge7       & Ge10 \\
 \hline

 Energy resolution (FWHM) at 84 keV     & 1.4 keV       & 1.3 keV   & 0.9 keV \\

 FWHM at 1332 keV                       & 2.3 keV       & 2.2 keV   & 1.9 keV \\

 Relative efficiency                    & 80\%          & 90\%      & 62\%  \\

 Ge crystal mass                        & 2096 g        & 1778 g    & 1040 g  \\

 Window material and thickness          & LB Cu 1.0 mm & HPAl 1.5 mm & HPAl 1.5 mm \\

 Top dead layer thickness               & 0.9 mm        & 0.3 $\mu$m    & 0.3 $\mu$m  \\

\hline
\multicolumn{4}{l}{LB Cu = Low Background Copper} \\
\multicolumn{4}{l}{HPAl = High Purity Aluminum} \\

\end{tabular}
\label{tab:detectors}
\end{center}
\end{table}

The Hf sample was stored 13 days underground before the low
background measurements to enable decay of short-lived cosmogenic
radionuclides. In the first measurement the hafnium sample was
installed directly on the endcap of the detector Ge10 (the
``set-up I'', see Fig.~\ref{fig:set-ups}). The measurements in the
set-up were continued over 40.4 days with the Ge10 detector and
36.4 days with the detector Ge7. The Ge10 detector is developed
for low-energy $\gamma$-rays measurements and has a very high
energy resolution and high detection efficiency to $\gamma$ quanta
in the energy region $\approx (50-80)$ keV where most of the
X-rays and $\gamma$ quanta expected in the two-neutrino mode of
the $2\varepsilon$ process to the ground state and to the lowest
excited level $2^+$ 76.5 keV of $^{174}$Yb should be emitted. The
Ge7 detector also has a rather high detection efficiency to
low-energy $\gamma$ quanta, despite a slightly worse energy
resolution.

After the first stage, the experiment was continued for 34.8 days
with the Ge6 detector instead of Ge10 (the second stage of the
experiment is named ``set-up II''). The Ge6 detector has a
comparatively high detection efficiency to middle and high-energy
$\gamma$ quanta, however its sensitivity to low-energy $\gamma$
quanta is substantially lower than that of the Ge7 and Ge10
detectors. Nevertheless, the detection efficiency of the detector
Ge6 is high enough to detect $\gamma$ quanta expected in the
$0\nu2\varepsilon$ and the $\varepsilon\beta^+$ processes with
energies $\sim(0.5-1)$ MeV (see Section \ref{sec:res}). Besides,
the detector was useful to estimate radioactive contamination of
the hafnium sample. The total exposure of the experiment was 42
g$\times$d for the isotope $^{174}$Hf\footnote{The exposure was
calculated as a product of the isotope mass on the sum of
measuring times of the four detectors used in the experiment.}.

\nopagebreak

\begin{figure}[htb]
\begin{center}
 \mbox{\epsfig{figure=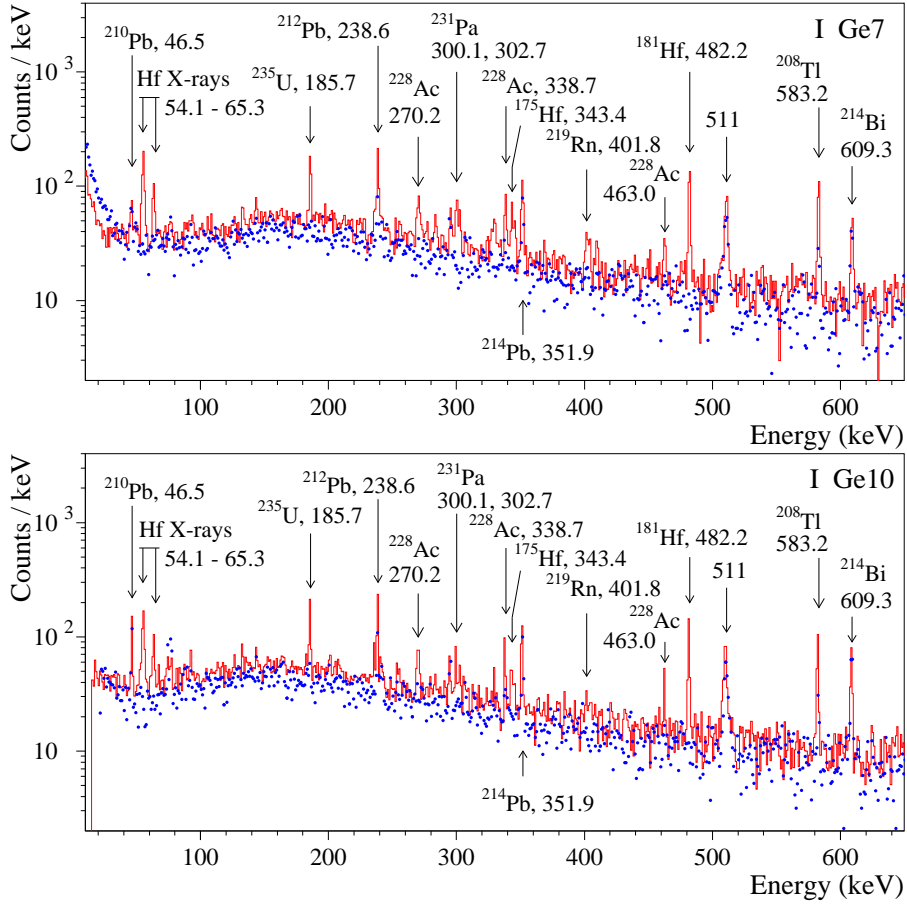,height=12.0cm}}
\caption{(Color online) Energy spectra accumulated with the Hf
sample (solid line) and without sample (dots) by
ultra-low-background HPGe $\gamma$ detectors Ge7 (over 38.4 d with
the hafnium sample and over 38.5 d without sample), and Ge10 (over
40.4 d with hafnium and 38.5 d background). The background energy
spectra are normalized to the times of measurements with the Hf
sample. Energy of $\gamma$ and X-ray quanta are in keV.}
 \label{fig:BGI}
 \end{center}
 \end{figure}

\begin{figure}[htb]
\begin{center}
 \mbox{\epsfig{figure=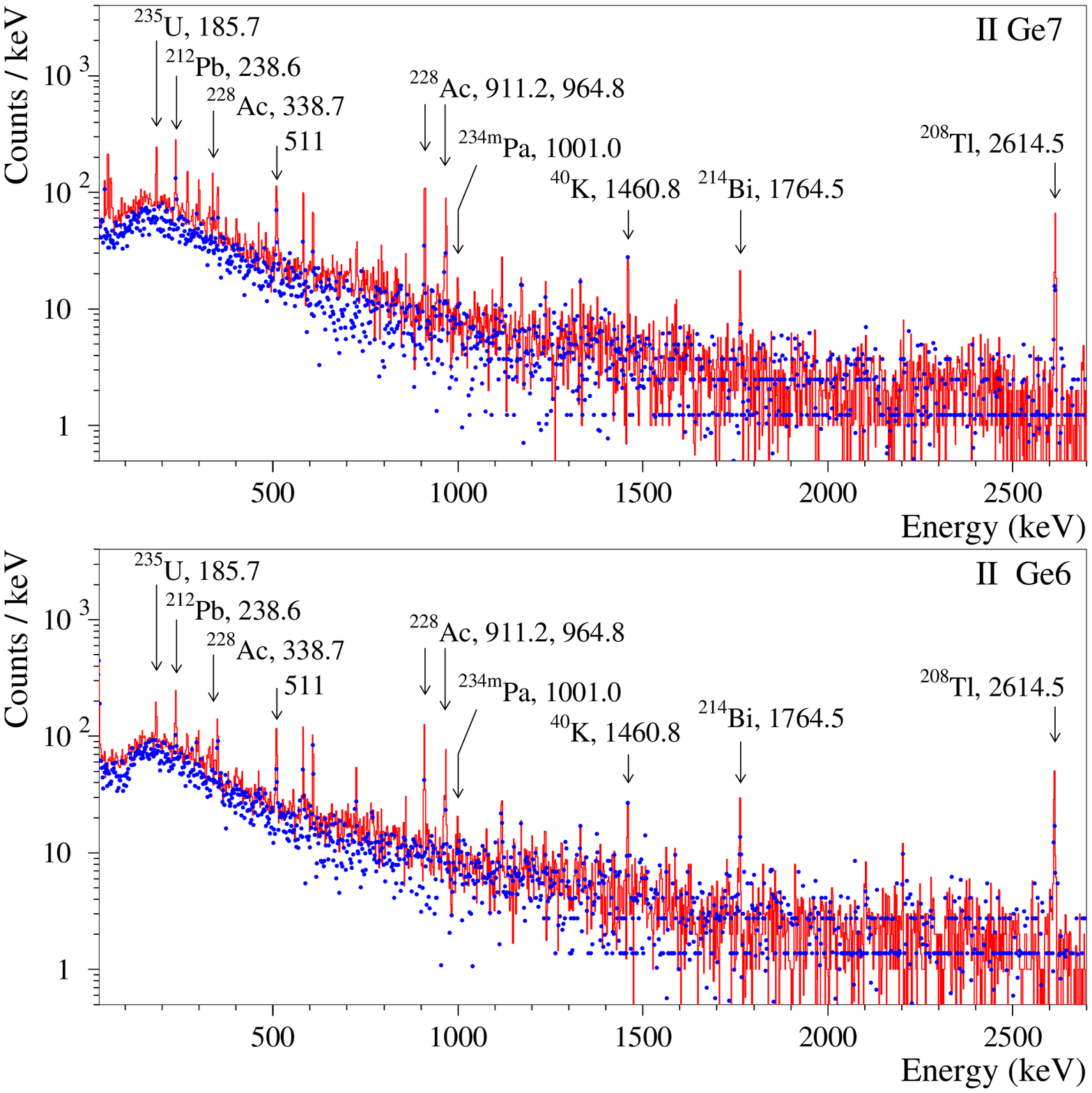,height=12.0cm}}
\caption{(Color online) Energy spectra accumulated with the
hafnium sample (solid line) and without sample (dots) by
ultra-low-background HPGe $\gamma$ detectors Ge7 (over 34.8 d with
the hafnium sample and over 28.1 d without sample), and Ge6 (over
34.8 d with hafnium and 25.3 d background). The background energy
spectra are normalized to the times of measurements with the Hf
sample. Energy of $\gamma$ quanta are in keV.}
 \label{fig:bgII}
 \end{center}
 \end{figure}

Energy spectra measured in the set-up I by the HPGe detectors Ge7
and Ge10 are shown in Fig.~\ref{fig:BGI}, while the spectra
accumulated in the set-up II are presented in Fig.~\ref{fig:bgII}.
Background spectra, normalized on the time of measurements with
the Hf sample, are shown in the Figures too.

\nopagebreak

\nopagebreak

There are many $\gamma$ peaks in the data that can be ascribed to
the naturally occurring primordial radionuclides: $^{40}$K,
daughters of the $^{232}$Th, $^{235}$U, and $^{238}$U families.
Some specific activities of hafnium radioactive nuclides were
observed too: $^{175}$Hf [electron capture with $Q_{EC}=683.9(20)$
keV and the half-life $T_{1/2}=70(2)$ days] and $^{181}$Hf [beta
active with $Q_{\beta}=1035.5(18)$ keV, $T_{1/2}=42.39(6)$ days].
It should be noted that the activities of the both radionuclides
in the sample decrease in time due to decay in the underground
conditions. We assume that the nuclides were generated by thermal
neutron captures by $^{174}$Hf and $^{180}$Hf, respectively (both
present in the Hf natural isotopic composition), and by
interactions with high energy cosmic neutrons on the Earth
surface, and especially, during the sample transportation by air.

Activities of the radionuclides in the hafnium sample were
calculated with the following formula:

\begin{equation}
A = (S_{sample}/t_{sample}-S_{bg}/t_{bg})/(\xi \cdot \eta)
 \end{equation}

\noindent where $S_{sample}$ ($S_{bg}$) is the area of a peak in
the sample (background) spectrum; $t_{sample}$ ($t_{bg}$) is the
time of the sample (background) measurement; $\xi$ is the
$\gamma$-ray yield for the corresponding transition
\cite{Firestone:1996}; $\eta$ is the full energy peak
detection efficiency. The efficiencies were Monte Carlo simulated
with the help of EGSnrc simulation package
\cite{Kawrakow:2017,Lutter:2018}.
The calculations were validated by comparison with the
experimental data obtained with $^{109}$Cd and $^{133}$Ba $\gamma$
sources (in the set-up I), and $^{109}$Cd, $^{133}$Ba, $^{134}$Cs,
$^{152}$Eu, $^{241}$Am $\gamma$ sources (set-up II). The standard
deviation of the relative difference between the Monte Carlo
simulations and the experiment is $(5-7)\%$ for $\gamma$ peaks in
the energy interval (53 -- 384) keV for the set-up I, and is 6\%
for $\gamma$ peaks in the energy interval (60 -- 1408) keV for the
set-up II. A summary of the estimated activities (limits) of
radioactive impurities in the Hf sample is given in
Table~\ref{tab:rad-cnt}.

\nopagebreak
\begin{table}[ht]
\caption{Radioactive contamination of the Hf sample measured by
HPGe $\gamma$-ray spectrometry. The activities of $^{175}$Hf and
$^{181}$Hf are given with reference date at the start of each
measurement for Set-up I and Set-up II (within brackets)
separately. Upper limits are given at 90\% C.L., the reported
uncertainties are the combined standard uncertainties.}
\begin{center}
\begin{tabular}{|l|l|l|}

 \hline
  Chain     & Nuclide       &  Activity in the sample (mBq) \\
 \hline
 ~          & $^{40}$K      & $\leq 1.4$ \\
 ~          & $^{60}$Co     & $\leq 0.11$ \\
 ~          & $^{137}$Cs    & $\leq 0.20$ \\
 ~          & $^{172}$Hf    & $\leq 3$ \\
 ~          & $^{175}$Hf    & $0.44\pm0.05$ $(0.18\pm0.04)$  \\
 ~          & $^{178m2}$Hf  & $\leq 0.06$ \\
 ~          & $^{181}$Hf    & $1.45\pm0.07$ $(0.12\pm0.04)$ \\
 ~          & $^{182}$Hf    & $\leq 0.5$ \\
 \hline
 $^{232}$Th & $^{228}$Ra    & $3.6\pm0.7$ \\

 ~          & $^{228}$Th    & $2.38\pm0.25$ \\
 \hline
 $^{235}$U  & $^{235}$U     & $3.8\pm0.5$  \\
 ~          & $^{231}$Pa    & $11\pm3$  \\
 ~          & $^{227}$Ac    & $2.0\pm0.5$ \\
 \hline
 $^{238}$U  & $^{234m}$Pa   & $11\pm5$ \\
  ~         & $^{226}$Ra    & $\leq0.7$  \\
 ~          & $^{210}$Pb    & $\leq50$ \\
  \hline

\end{tabular}
\label{tab:rad-cnt}
\end{center}
\end{table}

A peculiarity of the radioactive contamination is a significant
deviation of the $^{235}$U/$^{238}$U activities ratio in the Hf
sample\footnote{The observed ratio is $0.36(18)$, while the
expected one should be 0.046, assuming the natural isotopic
abundance of the uranium isotopes.}. The excess of $^{235}$U can
be explained by the application of gas centrifugation method to
remove zirconium in the hafnium production cycle (see Sec.
\ref{sec:exp}). Despite the details of the production process are
unknown, one could assume that the contamination by $^{235}$U
happened due to proximity between the industrial sites of the
centrifugation facilities to purify hafnium and to enrich uranium.
A more detailed discussion of the Hf sample radioactive
contamination one can find in \cite{Danevich:2019}.

\section{Results and discussion}
\label{sec:res}

No peculiarity was observed in the experimental data that could be
ascribed to the $2\beta$ decay processes in $^{174}$Hf. Thus, we
set limits on different modes and channels of the decay by using
the following formula:

\begin{equation}
\lim T_{1/2} = N \cdot \eta \cdot t \cdot \ln 2 / \lim S,
\end{equation}

\noindent where $N$ is the number of $^{174}$Hf nuclei in the
sample ($9.71\times10^{20}$), $\eta$ is the detection efficiency
for the effect searched for, $t$ is the measuring time, and $\lim
S$ is the number of events of the effect which can be excluded at
a given confidence level (C.L.). The detection efficiencies of the
detectors to the $\gamma$ (X-ray) quanta expected in different
modes and channels of the double beta processes in $^{174}$Hf were
simulated with the EGSnrc simulation package
\cite{Kawrakow:2017,Lutter:2018}, the decay events were generated
by the DECAY0 events generator \cite{DECAY0}.

\subsection{Search for double electron capture processes in $^{174}$Hf}

\begin{figure}[htb]
\begin{center}
 \mbox{\epsfig{figure=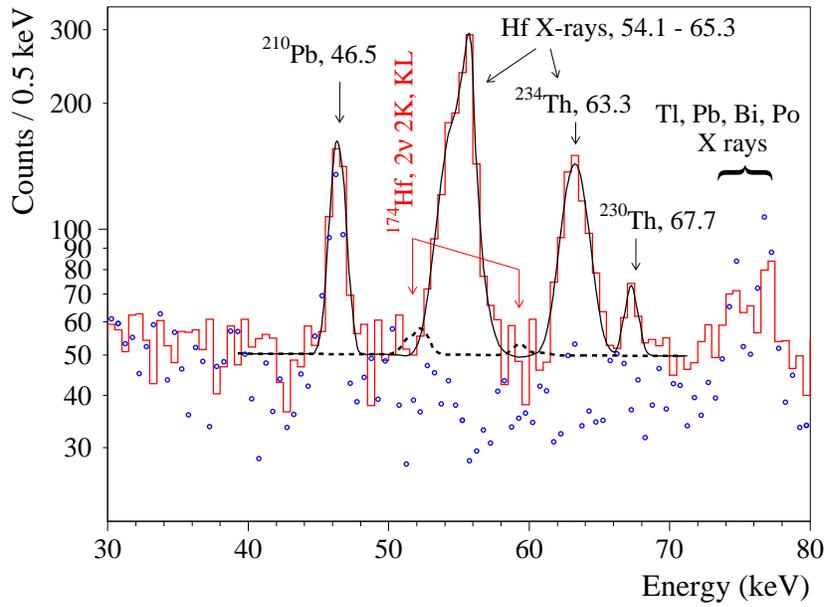,height=8.0cm}}
\caption{(Color online) The sum energy spectrum collected with the
detectors Ge7, Ge10 in the set-up I, plus the Ge7 detector in the
set-up II (solid histogram) in the energy region where $K$ X-ray
quanta are expected for the $2\nu 2K$ and $2\nu KL$ decays of
$^{174}$Hf. The fit of the data by the background model is shown
by solid line, while the excluded effect is presented by dashed
line. The background data accumulated with the detectors,
normalized on the measuring time, are shown by dots. Energy of
$\gamma$ and X-ray quanta are in keV.}
 \label{fig:2n2K}
 \end{center}
 \end{figure}

In case of the $2K$ and $KL$ capture in $^{174}$Hf, a cascade of
X-rays (and Auger electrons) of Yb atom with individual energies,
in particular, in the energy interval $(50.8-61.3)$ keV is
expected, while energies of the $2L$ capture X-ray quanta are
$\approx (7-10)$ keV, that are below the detectors' energy
thresholds. We took into account only the most intense X-rays of
ytterbium \cite{Firestone:1996}: 51.4 keV (the intensity of the
X-ray quanta is 27.2\%), 52.4 keV (47.4\%), 59.2 keV (5.2\%), 59.4
keV (10.0\%), and 61.0 keV (3.4\%). The energy spectra accumulated
with the Hf sample were fitted by the sum of several Gaussian
functions: five  peaks of $2\nu2K$ decay of $^{174}$Hf with
energies $(51.4-61.0)$ keV, a peak of $^{210}$Pb with energy 46.5
keV, Gaussian functions to describe the X-ray peaks of Hf, the
63.3 keV peak of $^{234}$Th, 67.7 keV peak of $^{230}$Th and a
straight line to describe the continuous background. A highest
sensitivity to the effect was achieved by analysis of a sum
spectrum of the detectors Ge7, Ge10 in the set-up I and of the Ge7
detector in the set-up II. The individual energy dependencies of
the detectors energy resolutions were taken into account in the
model. The data of the Ge6 detector was not used taking into
account a rather low detection efficiency of the detector to
low-energy $\gamma$ (X-ray) quanta. The best fit was achieved in
the energy interval ($39-71$) keV\footnote{It should be noted that
also two X-ray quanta with a total energy up to $\approx122$ keV
could be detected in coincidence in the $2\nu2K$ process, however,
the detection efficiency to the events is substantially lower than
that to one $K$ X-ray quanta.} with
$\chi^2/$n.d.f.$~=41.6/47=0.89$, where n.d.f. is number of degrees
of freedom. The fit gives an area of the effect $-7\pm26$ counts.
By using the recommendations \cite{Feldman:1998}, we took 36
counts as $\lim S$ with 90\% confidence level (C.L.)\footnote{All
the limits in the present work were set with 90\% C.L. by using
the recommendations \cite{Feldman:1998}.}. The sum energy spectrum
in the vicinity of the effect and the model of background are
presented in Fig.~\ref{fig:2n2K} together with the excluded
effect. It should be noted that here and below the estimations of
the $\lim S$ values include only the statistical errors coming
from the data fluctuations, and that systematic contributions have
not been considered. However, the statistical errors already do
include correlations to the background model. The detection
efficiency for the effect was calculated by the following formula:

\begin{equation}
\eta = \sum\eta_i\times t_i /\sum t_i,
\end{equation}

\noindent where $\eta_i$ are the individual detection efficiencies
and $t_i$ are the measuring times of the detectors used in the
analysis. The detection efficiency is estimated to be 1.24\% for
the whole X-rays distribution, and the following limit on the
$2\nu2K$ decay of $^{174}$Hf to the ground state of $^{174}$Yb was
set: $T^{2\nu2K}_{1/2}\geq 7.1\times10^{16}$ a. The detection
efficiency to the X-ray quanta in the energy interval
$(51.4-61.0)$ keV in the case of the $2\nu KL$ decay in $^{174}$Hf
to the ground state of $^{174}$Yb is lower (0.73\%) that results
in the following half-life limit: $T^{2\nu KL}_{1/2}\geq
4.2\times10^{16}$~a. The results are presented in Table
\ref{table:limits} together with the values of the $\lim S$ and
the detection efficiencies.

\nopagebreak

\clearpage

\begin{table}[ht]
\caption{The half-life limits on $2\varepsilon$ and
$\varepsilon\beta^+$ processes in $^{174}$Hf. The energies of the
$\gamma$ quanta ($E_{\gamma}$), which were used to set the
$T_{1/2}$ limits, are listed with their corresponding detection
efficiencies ($\eta$) and values of $\lim S$.}
\begin{center}
\begin{tabular}{lllllll}

\hline
 Channel                & Decay     & Level of  & $E_{\gamma}$  & $\eta$    & $\lim S$  & Experimental limit         \\
 of decay               & mode      & daughter  & (keV)         & (\%)      & (counts)  & $T_{1/2}$ (a) at 90\% C.L. \\
 ~                      & ~         & nucleus (keV)  & ~             & ~         & at 90\% C.L.   & ~           \\
 \hline

 $2K$                   & 2$\nu$    & g.s.      & $51.4-61.0$   & 1.24      & 36        & $\geq7.1\times10^{16}$ \\
 $KL$                   & 2$\nu$    & g.s.      & $51.4-61.0$   & 0.73      & 36        & $\geq4.2\times10^{16}$ \\
 $2K$                   & 2$\nu$    & $2^+$ 76.5& $51.4-61.0$   & 1.03      & 36        & $\geq5.9\times10^{16}$ \\
 $KL$                   & 2$\nu$    & $2^+$ 76.5& $51.4-61.0$   & 0.61      & 36        & $\geq3.5\times10^{16}$ \\
 $2L$                   & 2$\nu$    & $2^+$ 76.5& $76.5$        & 0.39      & 20.4      & $\geq3.9\times10^{16}$ \\

 $2K$                   & 0$\nu$    & g.s.      & 977.4         & 4.53      & 10.0      & $\geq5.8\times10^{17}$ \\
 $KL$                   & 0$\nu$    & g.s.      & 1028.9        & 4.46      & 3.0       & $\geq1.9\times10^{18}$ \\
 $2L$                   & 0$\nu$    & g.s.      & 1080.4        & 4.39      & 7.2       & $\geq7.8\times10^{17}$ \\

 $2K$                   & 0$\nu$    & $2^+$ 76.5& 900.9         & 4.67      & 8.4       & $\geq7.1\times10^{17}$ \\
 $KL$                   & 0$\nu$    & $2^+$ 76.5& 952.4         & 4.59      & 9.5       & $\geq6.2\times10^{17}$ \\
 $2L$                   & 0$\nu$    & $0^+$ 76.5& 1003.9        & 4.51      & 8.0       & $\geq7.2\times10^{17}$ \\

 $K\beta^+$             & (2$\nu$+0$\nu$) & g.s.    & 511       & 10.6      & 202        & $\geq1.4\times10^{17}$ \\
 $L\beta^+$             & (2$\nu$+0$\nu$) & g.s.    & 511       & 10.7      & 202       & $\geq1.4\times10^{17}$ \\
 \hline
 \end{tabular}
 \label{table:limits}
 \end{center}
 \end{table}

A similar group of Yb $K$ X-ray quanta is expected also in the
$2\nu 2K$ and $2\nu KL$ decays of $^{174}$Hf to the first $2^+$
76.5 keV excited level of $^{174}$Yb. The sensitivity with the
$(51.4-61.0)$ keV X-ray quanta analysis is higher than that with
the 76.5 keV $\gamma$ quanta. The reasons are: 1) a higher
detection efficiency to the X-ray quanta (in particular, because
in deexcitation of the 76.5 keV level mostly electrons are
emitted, conversion coefficient is equal 9.43 \cite{Browne:1999},
while the set-up is not sensitive to these electrons); 2) a rather
high counting rate in the vicinity of the energy 76.5 keV due to
the background caused by X-rays of Tl, Pb, Bi and Po. Thus, the
limits on the $2\nu 2K$ and $2\nu KL$ decays to the 76.5 keV level
were derived from the analysis of the energy region where the
group of $(51.4-61.0)$ keV X-ray quanta is expected (see
Fig.~\ref{fig:2n2K}). The detection efficiencies, excluded
effects' areas and obtained half-life limits are presented in
Table~\ref{table:limits}.

\begin{figure}[htb]
\begin{center}
 \mbox{\epsfig{figure=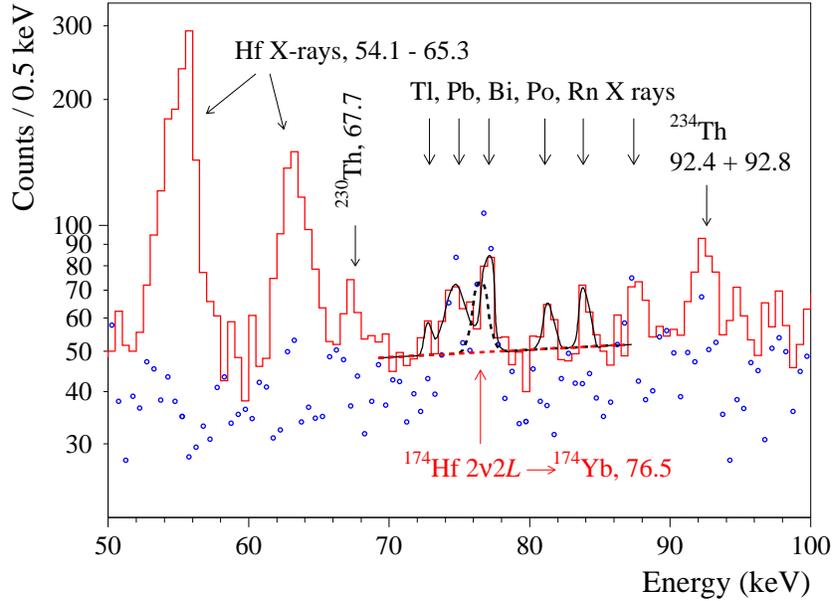,height=8.0cm}}
\caption{(Color online) Part of the sum energy spectrum collected
with the detectors Ge7, Ge10 in the set-up I, and with the Ge7
detector in the set-up II (solid histogram) in the energy region
where 76.5 keV $\gamma$ quanta emitted in the $2\nu2L$ decay of
$^{174}$Hf to the first $2^+$ 76.5 keV excited level of $^{174}$Yb
are expected together with the model of background (solid line).
The excluded 76.5 keV peak with area 20.4 counts is shown by
dashed line. The background energy spectrum accumulated with the
detectors, normalized on the measuring time, is shown by dots.
Energy of $\gamma$ and X-ray quanta are in keV.}
 \label{fig:2n2L-76}
 \end{center}
 \end{figure}

However, the limit for $2\nu2L$ process was obtained by analysis
of the 76.5 keV peak in the experimental data, taking into account
that $L$ X-ray quanta are below the acquisition energy threshold
of our experimental set-ups and cannot be detected in the present
experiment. The analysis of the data near 76.5 keV is rather
complicated, since there are many X-ray peaks in the energy
region. Thus, we have fitted the experimental sum spectrum of the
three detectors in the energy interval around the energy 76.5 keV
by sum of the most intensive $K$ X-ray quanta of Tl (72.9 keV), Pb
(75.0 keV), Bi (77.1 keV and 87.3 keV), Po (76.9 keV, 79.3 keV)
and Rn (81.1 keV, 83.8 keV). The model describes the experimental
data rather well with the $\chi^2/$n.d.f.$~=41.6/47=0.72$, giving
the 76.5 keV peak area $S=5.0 \pm 9.4$ counts, that corresponds to
$\lim S=20.4$ counts. The sum energy spectrum of the detectors
Ge7, Ge10 (set-up I) and of the detector Ge7 (set-up II) in the
energy region $50-100$ keV is presented in Fig.~\ref{fig:2n2L-76}
together with the background model and the excluded peak with
energy 76.5 keV. The obtained half-life limit on the $2\nu2L$
decay of $^{174}$Hf to the first $2^+$ 76.5 keV excited level of
$^{174}$Yb is given in Table~\ref{table:limits}.

\nopagebreak

\begin{figure}[htb]
\begin{center}
 \mbox{\epsfig{figure=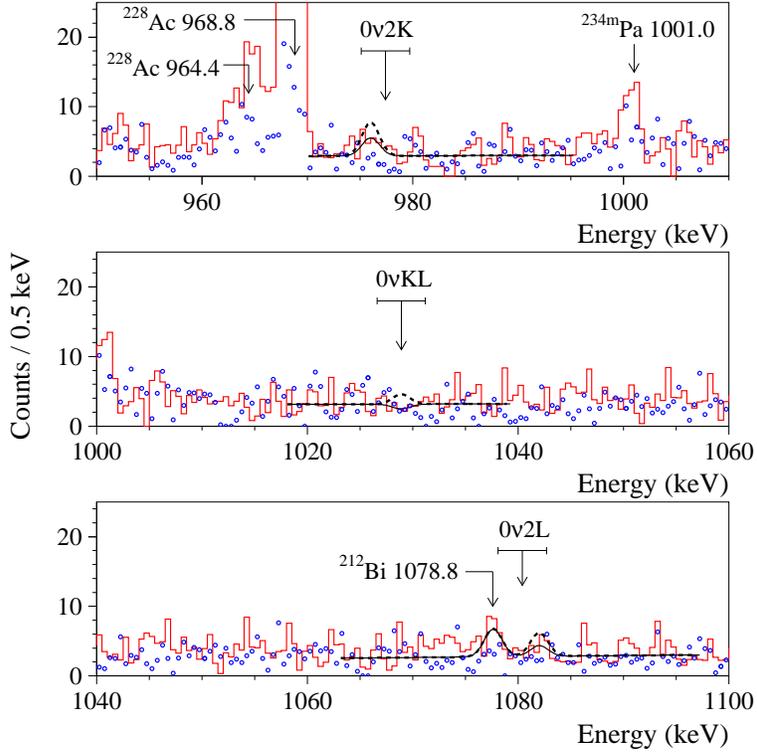,height=10.0cm}}
\caption{(Color online) Parts of the sum energy spectrum
accumulated with the Hf sample by the detectors Ge7 and Ge6 in the
set-up II, where the $\gamma$ peaks from the $0\nu2K$ (upper
panel), $0\nu KL$ (middle panel), and $0\nu2L$ (lower panel)
captures in $^{174}$Hf to the ground state of $^{174}$Yb are
expected. The fit of the data are shown by solid lines, while the
excluded peaks are presented by dashed lines. The horizontal lines
(above the arrows labelling the energy of the peaks searched for)
show the energy interval $\pm 2.3$ keV corresponding to the
uncertainty of the $Q_{2\beta}$ value of $^{174}$Hf. The
background data accumulated with the detectors, normalized on the
measuring time, are shown by dots. The energy of the background
$\gamma$ peaks are in keV.}
 \label{fig:0n2e}
 \end{center}
 \end{figure}

In case of the $0\nu$ double electron capture in $^{174}$Hf from
$K$ and $L$ shells to the ground state of the daughter nuclei, we
suppose that the energy excess in the process is taken away by
bremsstrahlung $\gamma$ quanta with an energy:
$E_{\gamma}=Q_{2\beta}-E_{b1}-E_{b2}$, where $E_{bi}$ are the
binding energies of the captured electrons on the $K$ and $L$
atomic shells of the daughter Yb nuclide. The energy spectrum
measured with the Hf sample by the detectors Ge7 and Ge6 in the
set-up II\footnote{The data gathered in the set-up I were not used
in the analysis due to the restricted at $\sim0.7$ MeV upper
energy thresholds of the detectors. The amplification was
increased with the intention to enable better peak-definitions in
the low-energy region of the spectrum by having more channels per
peak.} was fitted by a model constructed from a peak searched for
and a 1st degree polynomial function to describe the continuous
background (see Fig.~\ref{fig:0n2e}). In the case of the $0\nu 2L$
decay, the $\gamma$ peak of $^{212}$Bi with energy 1078.8 keV was
also included in the background model. The energy of the Gaussian
function used to describe the effect was varied taking into
account the uncertainty of the $Q_{2\beta}$ value ($\pm 2.3$ keV).
The fits to the energy spectrum provided ($5.4 \pm 2.8$) counts,
($-1.5\pm2.2$) counts, and ($3.3\pm2.4$) counts for the $0\nu 2K$,
$0\nu KL$, and $0\nu 2L$ peaks, respectively. The corresponding
$\lim S$ values are 10.0, 3.0 and 7.2 counts. The excluded peaks
are also shown in Fig.~\ref{fig:0n2e}.

\nopagebreak

\begin{figure}[htb]
\begin{center}
 \mbox{\epsfig{figure=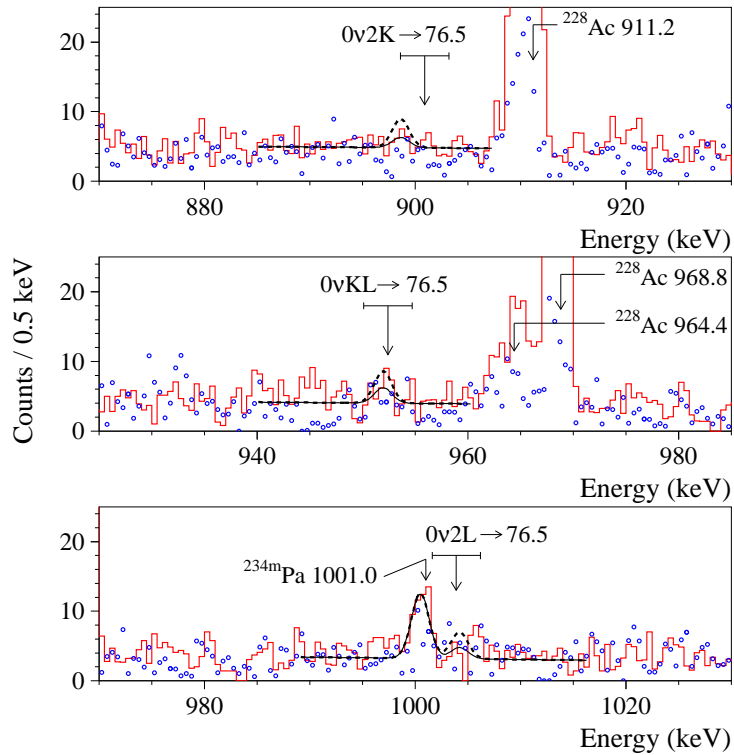,height=10.0cm}}
\caption{(Color online) Parts of the sum energy spectrum
accumulated with the Hf sample by the Ge7 and Ge6 detectors in the
set-up II, where the $\gamma$ peaks from the $0\nu2K$ (upper
panel), $0\nu KL$ (middle panel), and $0\nu2L$ (lower panel)
captures in $^{174}$Hf to the $2^+$ 76.5 keV excited level of
$^{174}$Yb are expected. The fits of the data are shown by solid
lines, while the excluded peaks are presented by dashed lines. The
horizontal lines (above the arrows labelling the energy of the
peaks searched for) show the energy interval $\pm 2.3$ keV
corresponding to the uncertainty of the $Q_{2\beta}$ value of
$^{174}$Hf. The background data accumulated with the detectors,
normalized on the measuring time, are shown by dots. The energy of
the background $\gamma$ quanta are in keV.}
 \label{fig:on2e-76}
 \end{center}
 \end{figure}

 \clearpage
A similar analysis was performed also for the $0\nu$ double
electron capture transitions to the $2^+$ 76.5 keV excited level
of $^{174}$Yb. The results of the analysis are shown in
Fig.~\ref{fig:on2e-76}. The obtained lower half-life limits for
the $0\nu2\varepsilon$ decays of $^{174}$Hf to the ground and the
$2^+$ 76.5 keV excited level of $^{174}$Yb are given in
Table~\ref{table:limits}.

\subsection{Search for electron capture with positron emission in $^{174}$Hf}

\begin{figure}[htb]
\begin{center}
 \mbox{\epsfig{figure=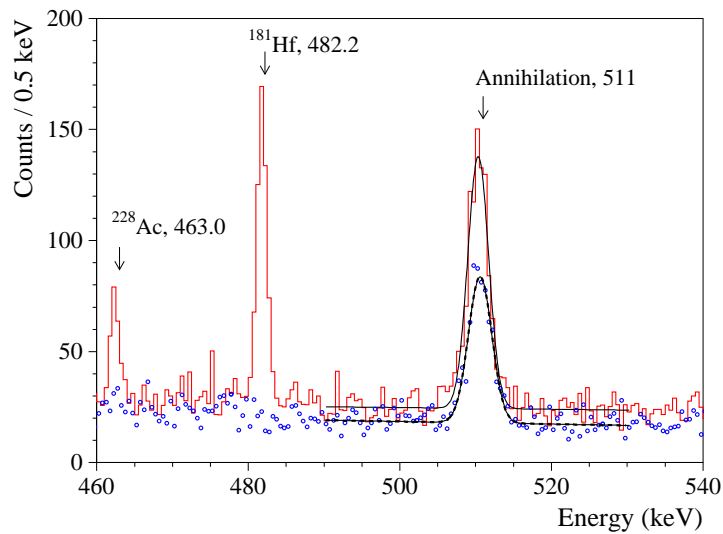,height=7.0cm}}
\caption{(Color online) Part of the energy spectra measured with
the Hf sample by the detectors Ge7, Ge10 (set-up I) and Ge7, Ge6
(set-up II) in the vicinity of the 511 keV annihilation peak. The
background data accumulated with the detectors, normalized on the
measuring time with the Hf sample, are shown by dots. The fits of
the data are shown by solid lines. The energy of the background
$\gamma$ peaks are in keV.}
 \label{fig:511}
 \end{center}
 \end{figure}

One positron with an energy up to $78 \pm 2.3$ keV should be
emitted in the $\varepsilon\beta^{+}$ decay of $^{174}$Hf. The
annihilation of the positron should give two 511 keV  $\gamma$'s
leading to an extra counting rate in the annihilation peak. The
sum of all four detectors' energy spectra was fitted in the energy
interval $(495-530)$~keV with a simple model constructed from a
511 keV peak (with a free parameter that describe the peak's
width) and a 1st degree polynomial function to describe
background. The fits of the experimental data in the vicinity of
the annihilation peak are shown in Fig.~\ref{fig:511}. There are
$783\pm33$ counts in the peak in the data accumulated with the
hafnium sample, and $508\pm32$ counts in the background data. The
Monte Carlo simulations show that the main part of the excess
($153\pm17$ counts) can be explained by decays of $^{228}$Ac and
$^{208}$Tl in the Hf sample. The residual peak area of $122\pm49$
counts, despite showing excess of more than two sigma, cannot be
accepted as effect of electron capture with positron emission in
$^{174}$Hf. The difference indicates presence of some systematics,
that needs more careful investigations, e.g., by using
isotopically enriched sample. Thus, assuming that the 511 keV peak
excess provides no evidence of the effect searched for, 202 counts
should be accepted as $\lim S$. Taking into account the detection
efficiency 10.6\% (10.7\%) for $K\beta^+$ ($L\beta^+$) decay mode,
we have obtained the same limit on the half-life of $^{174}$Hf
relatively to the electron capture from the $K$ and $L$ shells of
daughter atom with positron emission in $^{174}$Hf:
$T_{1/2}\geq1.4\times10^{17}$ a. The limits are presented in
Table~\ref{table:limits}. The limits are valid for both the $2\nu$
and $0\nu$ modes of the decay, since the modes cannot be
distinguished by the $\gamma$ spectrometry method.

\nopagebreak


\section{Conclusions}

A highly purified hafnium-disc with mass 179.8 g and the
dimensions $\oslash59.0\times 5.0$ mm was obtained by using the
double melting in vacuum by electron beam of a high purity sample
of metallic hafnium. The material was measured in the
HPGe-detector $\gamma$-ray spectrometry system located 225 m
underground at the HADES laboratory with an aim to search (at the
first time) for $2\varepsilon$ and $\varepsilon\beta^+$ decay of
$^{174}$Hf. No effect was observed after 75 days of data taking
but lower limits on the half-lives for the different channels and
modes of the decays were set on the level of $\lim T_{1/2}\sim
10^{16}-10^{18}$ a.

The sensitivity of the experiment could be advanced by using even
more highly purified from trace radioactive impurities hafnium
enriched in the isotope $^{174}$Hf, increasing the exposure, and
detection efficiency by application of thinner samples and
multi-crystal system of HPGe detectors. It should be stressed that
such an experiment looks practically realizable thanks to a
general possibility to apply gas centrifugation for Hf isotopical
enrichment, for the moment only the viable technology to produce
large enough amount of isotopically enriched materials.

\section{Acknowledgments}

This project received support from the EC-JRC open access project
EUFRAT under Horizon~2020. The group from the Institute for Nuclear
Research (Kyiv, Ukraine) was supported in part by the program of
the National Academy of Sciences of Ukraine ``Fundamental research
on high-energy physics and nuclear physics (international
cooperation)''. D.V.K. and O.G.P. were supported in part by the
project ``Investigations of rare nuclear processes'' of the
program of the National Academy of Sciences of Ukraine
``Laboratory of young scientists'' (Grant No. 0118U002328).


\begin{thebibliography}{99}
 \bibitem{Dolinski:2019} M.J.~Dolinski, A.W.P.~Poon, W.~Rodejohann, Neutrinoless Double-Beta Decay: Status and Prospects, Annu. Rev. Nucl. Part. Sci. 69 (2019) 219.
 \bibitem{Barea:2012} J.~Barea, J.~Kotila, F.~Iachello, Limits on Neutrino Masses from Neutrinoless Double-$\beta$ Decay, Phys. Rev. Lett. 109 (2012) 042501.
 \bibitem{Rodejohann:2012} W.~Rodejohann, Neutrino-less double $\beta$ decay and particle physics, J. Phys. G 39 (2012) 124008.
 \bibitem{Delloro:2016} S.~Dell'Oro, S.~Marcocci, M.~Viel, F.~Vissani, Neutrinoless Double Beta Decay: 2015 Review, AHEP 2016 (2016) 2162659.
 \bibitem{Vergados:2016} J.D. Vergados, H. Ejiri, F. Simkovic, Neutrinoless double beta decay and neutrino mass, Int. J. Mod. Phys. E 25 (2016) 1630007.
 \bibitem{Deppisch:2012} F.F.~Deppisch, M.~Hirsch, H.~P\"{a}s, Neutrinoless double-$\beta$ decay and physics beyond the standard model, J. Phys. G 39 (2012) 124007.
 \bibitem{Bilenky:2015} S.M.~Bilenky, C.~Giunti, Neutrinoless double-$\beta$ decay: A probe of physics beyond the standard model, Int. J. Mod. Phys. A 30 (2015) 1530001.
 \bibitem{Tretyak:2002} V.I.~Tretyak, Yu.G.~Zdesenko, Tables of double $\beta$ decay data -- an update, At. Data Nucl. Data Tables 80 (2002) 83.
 \bibitem{Elliott:2012} S.R.~Elliott, Recent progress in double beta decay, Mod. Phys. Lett. A 27 (2012) 123009.
 \bibitem{Giuliani:2012} A.~Giuliani, A.~Poves, Neutrinoless Double-Beta Decay, AHEP 2012 (2012) 857016.
 \bibitem{Saakyan:2013} R.~Saakyan, Two-Neutrino Double-Beta Decay, Annu. Rev. Nucl. Part. Sci. 63 (2013) 503.
 \bibitem{Cremonesi:2014} O.~Cremonesi, M.~Pavan, Challenges in Double Beta Decay, AHEP 2014 (2014) 951432.
 \bibitem{Gomes:2015} J.J.~G$\mathrm{\acute{o}}$mez-Cadenas, J.~Mart$\mathrm{\acute{i}}$n-Albo, Phenomenology of Neutrinoless Double Beta Decay, Proc. of Sci. (GSSI14) 004 (2015).
 \bibitem{Sarazin:2015} X.~Sarazin, Review of Double Beta Experiments, J. Phys.: Conf. Ser. 593 (2015) 012006.
 \bibitem{GERDA} M.~Agostini et al., Probing Majorana neutrinos with double-$\beta$ decay, Science 365 (2019) 1445.
 \bibitem{EXO-200} G.~Anton et al. (The EXO-200 Collaboration), Search for Neutrinoless Double-Beta Decay with the Complete EXO-200 Dataset, Phys. Rev. Lett. 123 (2019) 161802.
 \bibitem{CUORE} C.~Alduino et al. (CUORE Collaboration), First Results from CUORE: A Search for Lepton Number Violation via $0\nu\beta\beta$ Decay of $^{130}$Te, Phys. Rev. Lett. 120 (2018) 132501.
 \bibitem{NEMO-3} R. Arnold et al., Results of the search for neutrinoless double-$\beta$ decay in $^{100}$Mo with the NEMO-3 experiment, Phys. Rev. D 92 (2015) 072011.
 \bibitem{KamLAND-Zen} A.~Gando et al. (KamLAND-Zen Collaboration), Search for Majorana Neutrinos Near the Inverted Mass Hierarchy Region with KamLAND-Zen, Phys. Rev. Lett. 117 (2016) 082503.
 \bibitem{Maalampi:2013} J.~Maalampi, J.~Suhonen, Neutrinoless Double $\beta^+$/EC Decays, AHEP 2013 (2013) 505874.
 \bibitem{Blaum:2020} K.~Blaum et al., Double-Electron Capture, in preparation.
 \bibitem{Hirsch:1994} M.~Hirsch, K.~Muto, T.~Oda, H.V.~Klapdor-Kleingrothaus, Nuclear structure calculation of $\beta^+ \beta^+$, $\beta^+$/EC and EC/EC decay matrix elements, Z. Phys. A 347 (1994) 151.
 \bibitem{Wang:2017} M.~Wang et al., The AME2016 atomic mass evaluation, (II). Tables, graphs and references, Chin. Phys. C 41 (2017) 030003.
 \bibitem{Meija:2016} J.~Meija et al., Isotopic compositions of the elements 2013 (IUPAC Technical Report), Pure Appl. Chem. 88 (2016) 293.
 \bibitem{Danevich:2019} F.A.~Danevich et al., First search for $\alpha$ decays of naturally occurring Hf nuclides with emission of $\gamma$ quanta, arXiv:1910.02262v1 [nucl-ex], accepted to Eur. Phys. J. A.
 \bibitem{Wieslander:2009} J.S.E.~Wieslander et al., The Sandwich spectrometer for ultra low-level $\gamma$-ray spectrometry, Appl. Radiat. Isot. 67 (2009) 731.
 \bibitem{Hult:2013} M.~Hult et al., Comparison of background in underground HPGe-detectors in different lead shield configurations, Appl. Radiat. Isot. 81 (2013) 103.
 \bibitem{Firestone:1996} R.B.~Firestone et al., {\it Table of Isotopes}, 8th ed. (John Wiley, N.Y., 1996) and CD update (1998).
 \bibitem{Kawrakow:2017} I.~Kawrakow et al., The EGSnrc Code System: Monte Carlo simulation of electron and photon transport. Technical Report PIRS-701, National Research Council Canada (2017).
 \bibitem{Lutter:2018} G.~Lutter, M.~Hult, G.~Marissens, H.~Stroh, F.~Tzika, A gamma-ray spectrometry analysis software environment, Appl. Radiat. Isot. 134 (2018) 200.
 \bibitem{DECAY0} O.A.~Ponkratenko et al., Event Generator DECAY4 for Simulating Double-Beta Processes and Decays of Radioactive Nuclei, Phys. At. Nucl. 63 (2000) 1282.
 \bibitem{Feldman:1998} G.J.~Feldman, R.D.~Cousins, Unified approach to the classical statistical analysis of small signals, Phys. Rev. D 57 (1998) 3873.
 \bibitem{Browne:1999} E.~Browne, H.~Junde, Nuclear Data Sheets for A = 174, Nucl. Data Sheets 87 (1999) 15.
\end{thebibliography}
\end{document}